\documentclass[apj,twocolumn]{openjournal}
\usepackage{amsmath}
\usepackage{booktabs}
\usepackage{multirow}
\usepackage{color}
\usepackage{soul}

\usepackage[dvipsnames]{xcolor} %added later for color

\usepackage[breaklinks,colorlinks,citecolor=blue,urlcolor=blue]{hyperref}

\usepackage{orcidlink}

\renewcommand{\arraystretch}{1.2}  
\usepackage{listings}
\usepackage{color}
\definecolor{dkgreen}{rgb}{0,0.6,0}
\definecolor{gray}{rgb}{0.5,0.5,0.5}
\definecolor{mauve}{rgb}{0.58,0,0.82}
\definecolor{golden}{rgb}{0.86,0.65,0.01}
\begin{document}

% Use the \preprint command to place your local institutional report number 
% on the title page in preprint mode.
% Multiple \preprint commands are allowed.
%\preprint{}

\title{Asymmetric Drift Map of the Milky Way disk Populations between 8$-$16 kpc with LAMOST and Gaia datasets }

\author{Xin Li\,$^{1}$}
\author{Peng Yang\,$^{2}$}
\author{Hai-Feng Wang\,\orcidlink{0000-0001-8459-1036}$^{1,3,4}$}
\author{Qing Li\,$^{5}$}
\author{Yang-Ping Luo\,$^{1}$}
\author{Zhi-Quan Luo\,$^{1}$}
\author{Guan-Yu Wang\,$^{1}$}

\affiliation{$^1$Department of Astronomy, China West Normal University, Nanchong, 637002, P.\,R.\,China}
\affiliation{$^2$Urban Vocational College of Sichuan, Chengdu, 610101, P.\,R.\,China}
\affiliation{$^3$CREF, Centro Ricerche Enrico Fermi, Via Panisperna 89A, I-00184, Roma, Italy}
\affiliation{$^4$Department of Physics, Sapienza, University of Rome, Piazzale Aldo Moro 5, I-00185 Rome, Italy}
\affiliation{$^5$Jiangmen Experimental Middle School of Guangdong, Jiangmen, 529000, P.\,R.\,China}

\email{Corresponding author: hfwang@bao.ac.cn}
 
\begin{abstract}

The application of asymmetric drift (AD) tomography across different populations provides valuable insights into the kinematics, dynamics, and rotation curves of the Galactic disk. By leveraging common stars identified in both the LAMOST and Gaia surveys, alongside Gaia DR3's circular velocity curve, we conducted a qualitative exploration of asymmetric drift distributions within the Galactic disk spanning distances from 8 to 16 kpc. In the R-Z plane, we observed that the asymmetric drift is minimal near the mid-plane of the Galactic disk and gradually increases with vertical distance, resulting in a distinctive ``horn" shape. Additionally, our analysis revealed that populations with higher [$\alpha$/Fe] ratios exhibit greater asymmetric drift compared to those with lower [$\alpha$/Fe] ratios. Specifically, we found the asymmetric drift around the solar location to be approximately 6 km s$^{-1}$, with a median value of 16 km s$^{-1}$ across the entire sample. Notably, the median asymmetric drift in the northern region of the Galactic disk (20 km s$^{-1}$) surpasses that in the southern region (13 km s$^{-1}$), with errors remaining within 2 km s$^{-1}$. Furthermore, our investigation into mono-age stellar populations unveiled that older stellar populations tend to exhibit larger asymmetric drift and velocity dispersion, aligning closely with predictions from previous numerical models. Finally, based on chemical compositions, we observed that the median asymmetric drift of the thick disk significantly exceeds that of the thin disk and found that star formation within the thick disk primarily occurred earlier than 8-10 billion years, whereas the thin disk's predominant star formation period spanned 6-8 billion years ago.

\keywords{Milky Way Galaxy; Milky Way disk;}

\end{abstract}

\section{Introduction}

The Milky Way serves as a pivotal case study for comprehending the structure, formation, and evolution of disk galaxies. Its disk encompasses a significant proportion of baryonic matter characterized by ongoing star formation and secular evolution \citep{vanderKruit2011}. By scrutinizing the six-dimensional (6D) phase space distribution and dynamic characteristics, we can unravel the underlying physical mechanisms governing both internal interactions within the Galactic disk and external perturbations. This approach also facilitates the study of Galactoseismology, which aims to discern the oscillatory behavior and resonant phenomena within the Milky Way, e.g., \citet{Widrow2012,Widrow2014,Carlin2013,Williams2013,Siebert2012,Lopez2020,2018A&A...612L...8L,wang2018a,wang2018b,wang2019,wang2020a,wang2020b,wang2020c,wang2022,wang2023a, wang2023b,Khoperskov2022,lixiang2022,Yang2023,Drimmel2022,2023A&A...673A.115A}. 

\citet{Antoja2018} utilized stellar velocity distribution data from Gaia DR2 to identify substructures such as arches, ridges, and snail shells in the solar neighborhood, whose origins remain insufficiently understood due to unclear physical mechanisms. Prior research suggests that these substructures may stem from internal perturbations induced by the outer Lindblad resonance of the Galactic bar and the dynamics of spiral arms \citep{Kawata2018}, as well as stellar orbits influenced by spiral resonance \citep{Barros2020}. Additionally, the interaction between the Milky Way and the Sagittarius galaxy plays a significant role in shaping the Galactic disk \citep{Binney2018,Khanna2019,Laporte2020}, which is crucial for comprehending phenomena such as flaring and the non-equilibrium state of the disk \citep{2019MNRAS.483.3119T,2020NatAs...4..965R,Yu2021}, among others.

When investigating the three-dimensional velocity distribution of stars, it becomes evident that stars demonstrate a non-axisymmetric distribution of azimuthal velocity, a phenomenon known as asymmetric drift. Traditionally, stellar motion in the Galactic disk has been conceptualized as relatively simple circular or elliptical orbits. However, due to perturbations or the non-equilibrium state of the disk, certain stars deviate from the plane and exhibit upward motion, resulting in a reduction in angular momentum. Consequently, the observed phenomenon of asymmetric azimuthal velocity distribution is termed asymmetric drift \citep{Binney2008}. Alternatively, asymmetric drift can be defined as the difference between the velocities of stars following perfectly circular orbits in the Milky Way and the median tangential or azimuthal velocity of the stellar population \citep{Golubov2013}.

\citet{Golubov2013} elucidated the correlation between age and velocity dispersion for both V$a$ (asymmetric drift value) and V$_{\odot}$ (solar peculiar motions). This insight stems from applying the Jeans equation to stellar populations of varying ages, wherein the velocity dispersion of the sample progressively increases with age, reflecting the age-velocity dispersion relations. Addressing this challenge necessitates utilizing stellar populations that are not extremely young, as the median tangential velocity linearly depends on the square of the velocity dispersion. This characteristic enables extrapolation to zero velocity dispersion, thereby facilitating the determination of V$_{\odot}$.

In \citet{Golubov2013}, the Jeans equation adopts the form of Eq. 4, known as the linear Strömgren equation. They recalculated the Local Standard of Rest (LSR) based on standard assumptions of radial scale lengths independent of velocity dispersion in each metallicity bin, resulting in a new determination of the solar peculiar motion. Additionally, \citet{Dehnen1998} measured the velocity of the Sun relative to the LSR in the direction of Galactic rotation and found that the asymmetric drift linearly depends on the square of the total velocity dispersion of the stellar population. Utilizing Hipparcos data, \citet{Aumer2009} employed a similar approach to obtain a solar velocity estimate with a small error margin for V$_{\odot}$. Various methods and models appear to influence the calculation of the Sun's peculiar motion \citep{Binney2010}, and modeling progress regarding asymmetric drift around the Sun can be found in \citet{2014MNRAS.439.1231B}. The local rotation curve and asymmetric drift were thoroughly discussed and analyzed using the RAVE and SEGUE samples by \citet{Sysoliatina2018}. More recently, there has been a focus on considering the asymmetric drift effect in the rotation curve in greater detail \citep{2023arXiv230605461J}, utilizing the Jeans equation. Subsequently, \citet{2023A&A...676A.134P} determined the rotation curve and asymmetric drift within 14 kpc using Bayesian estimation and the Gaia DR3 dataset.

\citet{Williams2013} discovered that the asymmetric drift varies significantly with vertical height, with a potential difference of up to 40 km/s between the midplane and Z = 2 kpc. Investigating disk kinematics, \citet{Robin2017} utilized data from Gaia DR1 and the Radial Velocity Measurement Experiments (RAVE4) high latitude sample \citep{Kordopatis2013,Steinmetz2020a,Steinmetz2020b}. They also delved into the asymmetric drift of the Galactic disk by fitting the stackel potential to orbits. Employing this approach, \citet{Robin2017} explored the asymmetric drift of the thin disk and thick disk (young and old) across different ages as a function of Galactocentric distance and vertical distance, based on mock data. In the $R_{\text{lag}}$-V$_{\text{lag}}$ diagram, they observed that the asymmetric drift tends to decrease as the distance from the Galactic center increases. Conversely, in the $Z_{\text{lag}}$-$V_{\text{lag}}$ diagram, the asymmetric drift gradually increases with vertical distance.

The contribution of asymmetric drift extends beyond the stellar velocity profile to influence the rotation curve of the Milky Way. However, our understanding of the asymmetric drift distribution across the Galactic disk in different age and abundance populations beyond the solar neighborhood remains limited. This paper aims to address this gap by leveraging data from LAMOST DR4 and Gaia DR3, coupled with a fundamental dynamic formula, to quantify the asymmetric drift of mono-age-abundance populations of stars within the Galactic disk. The objective is to provide observational maps of asymmetric drift to the scientific community. These maps will serve as valuable resources for future discussions on topics such as disk asymmetries and rotation curves, offering both statistical insights and qualitative perspectives. Unlike previous works that relied on the Jeans analysis, this study opts to fully exploit the abundant phase space properties present in different populations under the assumption of the disk rotation curve. The subsequent step will involve showcasing rotation curve patterns in different populations.

The paper is structured as follows: Section 2 describes the data sources, including information on chemical abundance, age distribution, and three-dimensional velocities of the stellar sample. Additionally, outlines the model or formula used to calculate asymmetric drift. Section 3 presents the main findings of the study, including the asymmetric drift maps in different parameter spaces. Section 4 discusses the implications and interpretations of the results obtained for asymmetric drift. Section 5 summarizes the key findings of the paper and discusses their significance in the context of understanding disk asymmetries and rotation curves in the Galactic disk.

\section{Data and Methods}  
\subsection{Sample}

The Guo Shoujing Telescope, also known as the Large Sky Area Multi-Object Fiber Spectroscopic Telescope (LAMOST), represents a novel telescope design featuring a large field of view and aperture, categorized as a ``reflective Schmidt telescope." With an effective aperture of 4 meters, LAMOST leverages controllable fiber positioning technology to accommodate 4,000 optical fibers within a focal plane measuring 1.75 meters in diameter. Its expansive field of view spans five degrees, while its spectral coverage ranges from 370 nm to 900 nm \citep{Zhao2012, Deng2012}. This wide spectral range enables the acquisition of spectral data across various wavelengths, facilitating investigations into stellar age, chemical abundance, and other physical parameters \citep[e.g.,][]{2020MNRAS.495.1252Z,2023arXiv231017196W}.

The stellar parameters utilized in our sample are derived from the red giant branch stars (RGB) of LAMOST DR4, as selected by \citet{Wu2019}. The age and its associated uncertainty are determined using the kernel principal component analysis method (KPCA), with an estimated uncertainty of 30\%. The average uncertainty of the radial velocity is reported as 5 km/s, obtained through the LAMOST stellar parameter pipeline of Peking University (LSP3). When employing LSP3 to determine stellar parameters, the uncertainty of effective temperature (Teff) is stated as 100 K, while the uncertainty of surface gravity (log (g)) is reported as 0.1 dex \citep{Xiang2017}. Metallicity and elemental abundances are obtained through cross-matching with the stellar catalogue derived by \citet{Xiang2019} using the data-driven Payne method. \citet{Xiang2019} utilized the data-driven Payne (DD-Payne; \citet{Ting2019}) modeling method to derive parameters and element abundances for 6 million stars from the low-resolution ($R\mbox{-}1800$) spectra of LAMOST DR5. For the adopted sample, the reported error of metallicity is 0.1 dex, while the error of [$\alpha$/Fe] is approximately 0.05 dex. The distance estimates in the catalog are derived using the Bayesian estimation method, as calibrated by \citet{Xu2020}, with an estimated uncertainty of about 15\%. 

The proper motion data utilized in this paper is sourced from the Gaia DR3 dataset, which we cross-match with our sample. Notably, we do not impose a cut on the renormalized unit weight error (RUWE) in order to maintain sufficient sampling. While this decision may increase dispersion, it does not alter the final conclusions drawn in this specific study. The Gaia DR3 dataset provides comprehensive information including the positions of stars, proper motion, parallax, and other relevant parameters. It is noteworthy that the proper motion and parallax measurements in Gaia DR3 exhibit high accuracy, as acknowledged in \citet{Gaia2022}. For additional details, interested readers may refer to \citet{Yang2023}.

\subsection{Velocity coordinates}
The kinematic distributions of the red giant branch (RGB) stars are derived through coordinate transformations using Galpy \citep{Bovy2015}. For the parameters of the coordinate transformation, we adopt the following: The distance from the position of the Sun to the center of the Milky Way is 8.34 kpc \citep{Reid2014}. The vertical distance from the position of the Sun to the Galactic disk mid-plane is 27 pc \citep{Chen2001}. For the local standard of rest (LSR), we use the value of 238 km/s calculated by \citet{Schonrich2012}.
The solar peculiar motions relative to the LSR are (9.58, 10.5, 7.01) km/s, respectively \citep{Tian2015}. Combining these parameters, we calculate the six-dimensional properties for 300,433 RGB stars. It is important to note that different choices of solar motions will not alter the final conclusions drawn in this study. We adopt Galactocentric Cartesian coordinates with the following conventions: $X$: Increasing outward from the Galactic center. $Y$: In the direction of rotation. $Z$: Positive towards the North Galactic Pole (NGP).
Cylindrical velocities $V_R$, $V_\phi$, and $V_Z$ are defined as positive with increasing $R$, $\phi$, and $Z$, respectively.

\begin{figure}
  \centering
  \includegraphics[width=0.4\textwidth,height=0.5\textwidth]{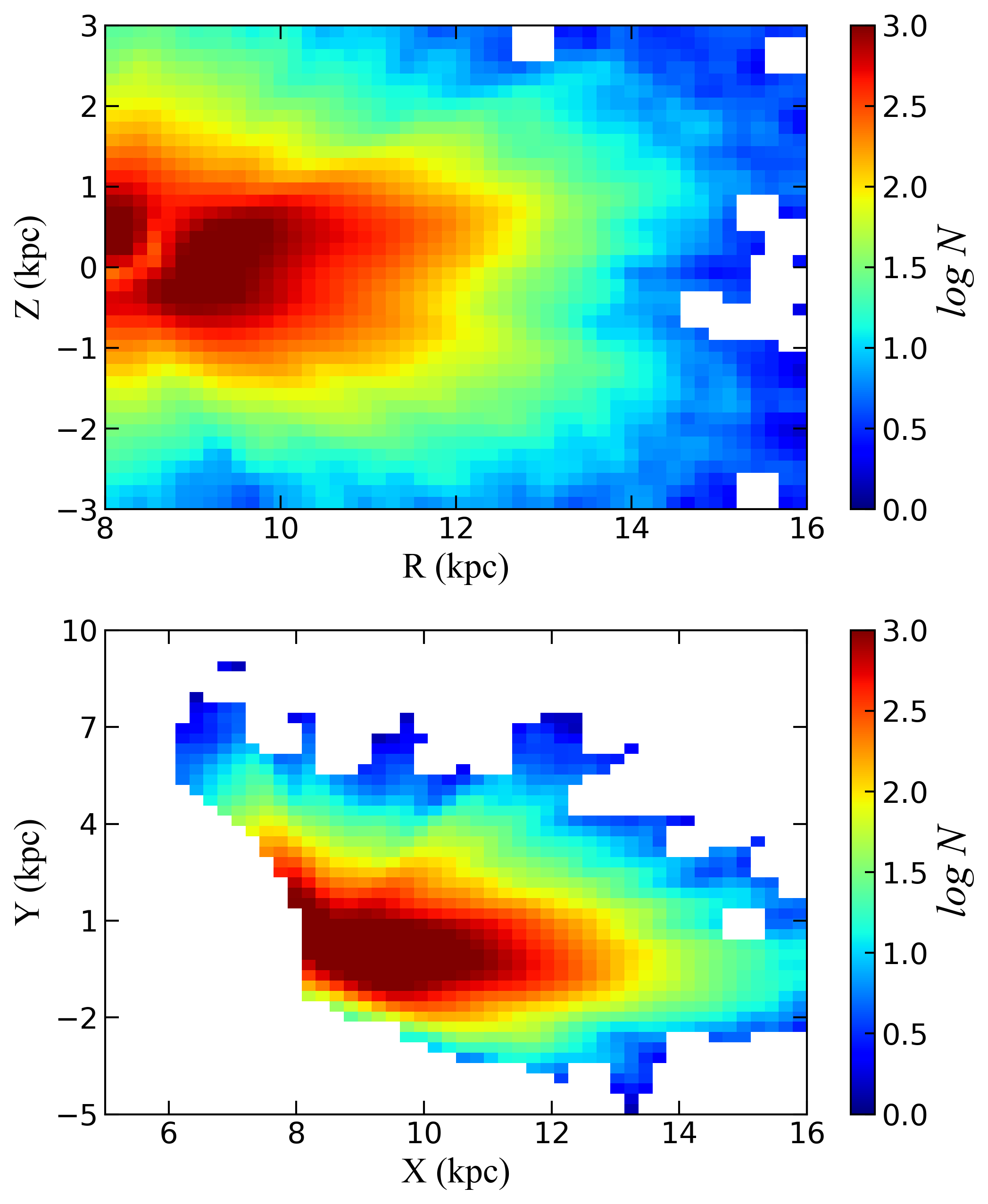}
  \caption{Top panel: The smoothed distribution of RGB stars in the $R-Z$ plane utilized in this study reveals a logarithmic scale of star counts, indicating a higher density of stars on the northern side of the Galactic disk compared to the southern side. Bottom panel: The distribution of RGB stars in the $X-Y$ plane is depicted, with bin sizes set at 0.16 kpc for R, 0.15 kpc for Z, 0.22 kpc for X, and 0.32 kpc for Y. Each pixel represents a minimum of 1 star count, with a smoothing technique applied for visualization.}
  \label{X-X-COUNTS}
\clearpage
\end{figure}

We make some constraints as follows in order to mainly focus on the disk region:

(1) 8 \textless R \textless 16 kpc and $-$3 \textless Z \textless 3 kpc (we focus on the outer thin and thick disk region);

(2) 0 \textless Age \textless 14 Gyr (removing some stars older than 14 Gyr, which is not true);

(3) [Fe/H] \textgreater $-$1.5 dex and [$\alpha$/Fe] \textless 0.35 dex; 

(4) $-$150 \textless V$_R$ \textless 150 km s$^{-1}$, $-$50 \textless V$_\phi$ \textless 350 km s$^{-1}$ and $-$150 \textless V$_Z$ \textless 150 km s$^{-1}$;

After applying these constraints, we ultimately obtain a sample of 231,874 RGB stars. Figure \ref{X-X-COUNTS} illustrates the spatial distribution of our stellar sample on both the cylindrical ($R-Z$) and Cartesian ($X-Y$) coordinate planes. The coloration represents the number of stars on a logarithmic scale. Examining the density distribution of stars, we observe a higher concentration of stars on the northern side of the disk compared to the southern side in our sample. Additionally, indications of disk flaring can be discerned in the top panel of the figure, although it is important to note that the influence of selection effects cannot be entirely ruled out.

\subsection{Method for Asymmetric drift}  

The method for calculating the asymmetric drift, as adopted from \citet{Golubov2013} and elaborated further in \citet{Binney2008} (Section 4), involves the following equation:

\begin{equation}\label{model1}
V_{a} = \bar V_{c(R)} - \bar V_{\phi(R)}=\triangle V - V_{\sun}
\end{equation}

\begin{equation}\label{model2}
V_{c(R)} = 229-2.3 (R-R_{\sun}) 
\end{equation}

\begin{figure*}
  \centering  \includegraphics[width=0.95\textwidth,height=0.5\textwidth]{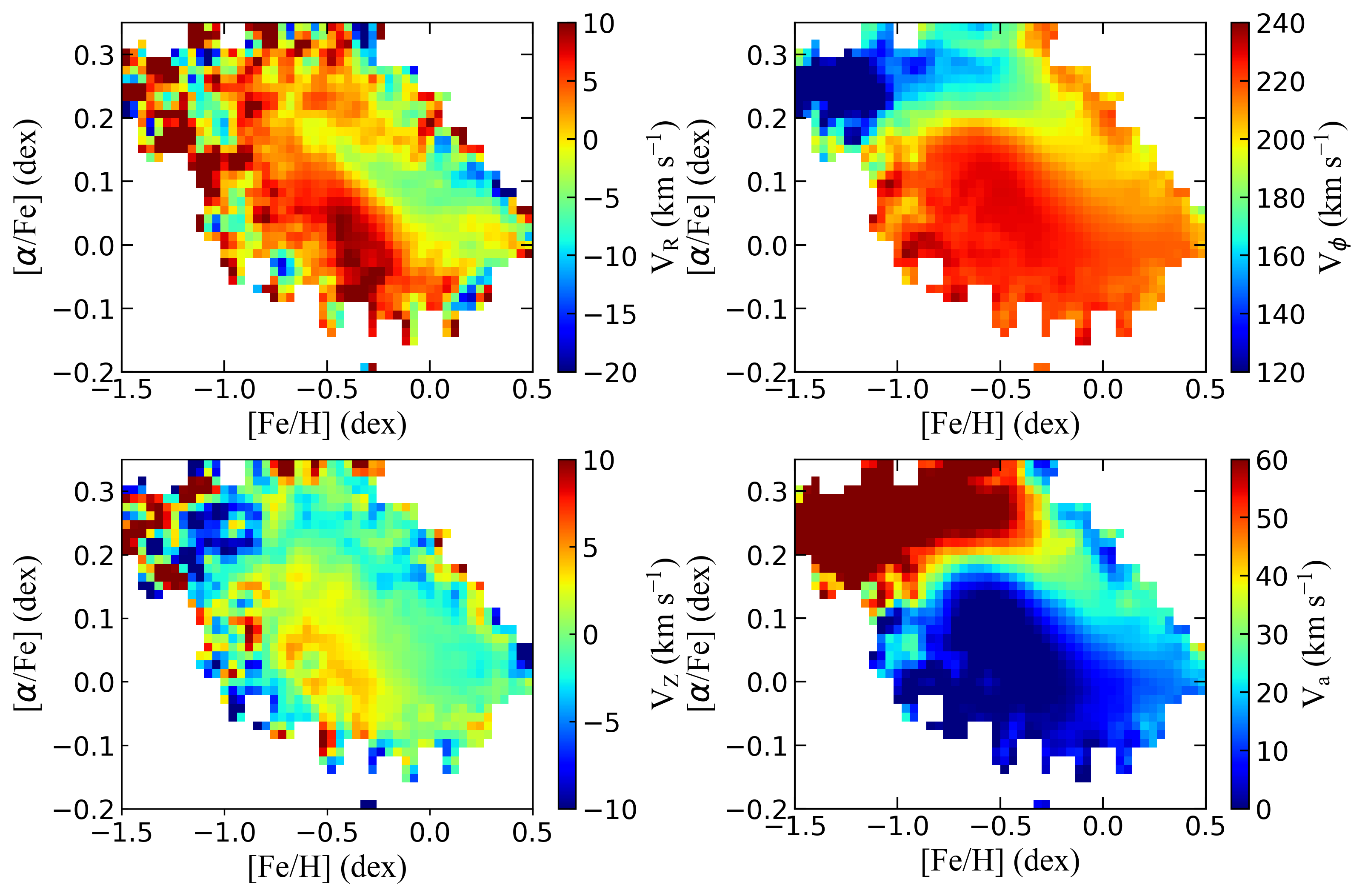}
  \caption{Top panel: The stars' Galactic radial (V$_{R}$) and azimuthal (V$_{\phi}$) velocities are depicted on the [Fe/H] and [$\alpha$/Fe] plane. Bottom panel: The stars' Galactic vertical (V$_{Z}$) and asymmetric drift (V$_{a}$) velocities are illustrated on the [Fe/H] and [$\alpha$/Fe] planes. The bin size is set at 0.04 dex for [$\alpha$/Fe] and 0.01 dex for [Fe/H]. Throughout this study, we ensured a minimum of 1 star in each pixel using a smoothing technique, with all values represented as medians, which align closely with the mean values in this analysis.}
  \label{abundance-X}
\end{figure*}

Where $R$ denotes the distance from the Galactic center, with the Sun positioned 8.34 kpc away from it. $\bar{V}_c(R)$ signifies the median rotational velocity, while $\bar{V}_\phi(R)$ represents the median azimuth velocity within the volume. The first method for calculating $V_a$ involves subtracting the median azimuthal velocity from the median rotational velocity. Alternatively, the latter formula calculates $V_a$ as the median tangential velocity minus the peculiar motion of the Sun, primarily applicable to the local volume. Our calculations of the asymmetric drift are based on Equation (\ref{model2}), which assumes a specific rotation curve profile. While alternative assumptions may influence the absolute values of the asymmetric drift, the overall conclusions remain robust. The circular velocity profile and its slope, drawn from \citet{Eilers2019} and \citet{wang2023a} respectively, demonstrate consistent trends despite slight variations in solar motions, local standard of rest (LSR) values, and stellar tracers among different rotation curve models. As emphasized, our focus is on discerning statistical laws and patterns across different populations. Future work will delve into rotation curves tailored to specific population dynamics.

\section{Results}
\subsection{Asymmetric Drift Distribution for the Entire Sample}

We determine the asymmetric drift of the entire sample and illustrate its distribution across the Galactic disk using the equation outlined by \citet{Golubov2013}. Figure~\ref{abundance-X} showcases the velocity distribution and asymmetric drift of the sample in chemical space ([Fe/H]-[$\alpha$/Fe]). Notably, while there is no discernible trend in the radial and vertical velocity distributions, the azimuthal velocity reveals a noteworthy pattern. Specifically, the decrease of [$\alpha$/Fe] with azimuthal velocity suggests the dominance of different components of the Milky Way at varying velocities (thin disk, thick disk, and halo). The velocities of stars in Galactic coordinates exhibit distinct patterns: the thin disk, characterized by [$\alpha$/Fe] around 0.0 dex, rotates faster than the thick disk, which typically exhibits [$\alpha$/Fe] around 0.2 dex. Moreover, the bottom-right panel clearly demonstrates that while the thin disk population maintains a velocity distribution within 10 km/s, the thick disk can extend beyond 40 km/s.

\begin{figure*}
  \centering
  \includegraphics[width=0.8\textwidth,height=0.4\textwidth]{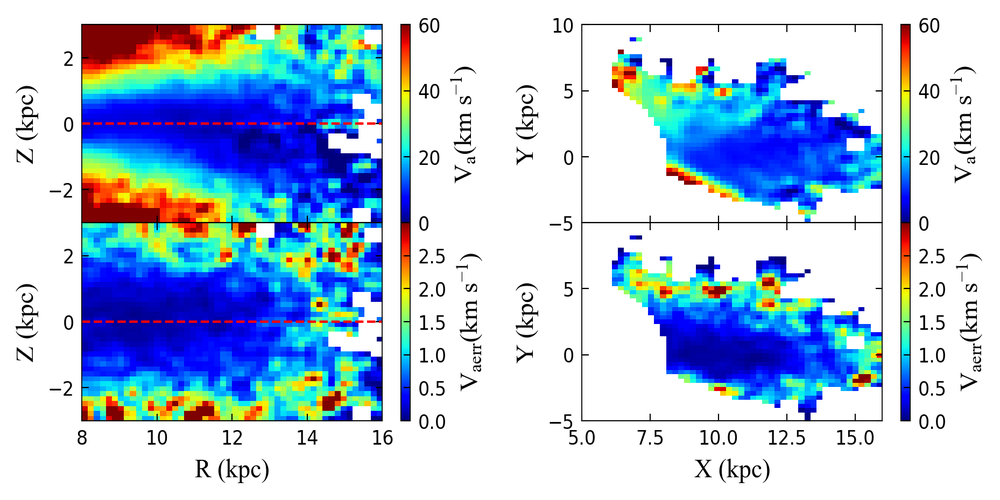}
  \caption{The asymmetric drift distribution of the final sample is depicted. The top diagrams illustrate the distribution of asymmetric drift in the $R-Z$ and $X-Y$ planes, while the bottom diagrams display the corresponding errors obtained through Bootstrap. The vertical range of the right panel corresponds to the height range of the left one, with the horizontal red dotted line denoting Z=0.}
  \label{R-Z-VA}
\clearpage
\end{figure*}

Figure~\ref{R-Z-VA} illustrates the distribution of asymmetric drift and its associated errors across $R-Z$ and $X-Y$ phase space. In the upper left panel, the asymmetric drift spans from 0 to 60 km/s, with a similar pattern observed in the upper right panel. The lower panels depict corresponding error distributions obtained through bootstrapping, where uncertainties primarily stem from distance, proper motion, and line-of-sight velocity uncertainties. Bootstrap errors are determined by resampling (with replacement) 100 times for each bin, with uncertainties calculated using the 15th and 85th percentiles of the sample. For both components, the majority of error values are less than 2 km/s for the asymmetric drift. Utilizing Equation (\ref{model2}), we find that the median value of the asymmetric drift for the entire sample is $V_a = 16$ km/s in the $R-Z$ plane. Analyzing the distribution of asymmetric drift in the $R-Z$ plane, we observe that near the Galactic disk's plane, the asymmetric drift exhibits a smaller magnitude on average. However, as the vertical distance increases, the asymmetric drift gradually rises, resembling a ``horn" shape in the $R-Z$ plane. This trend aligns closely with the findings of \citet{Katz}. Comparing the distribution of asymmetric drift in cylindrical and Cartesian coordinates, we observe no clear trend in the latter. However, it is notable that the asymmetric drift appears small around $Y=0$.

\begin{figure}
  \centering
  \includegraphics[width=0.48\textwidth,height=0.3\textwidth]{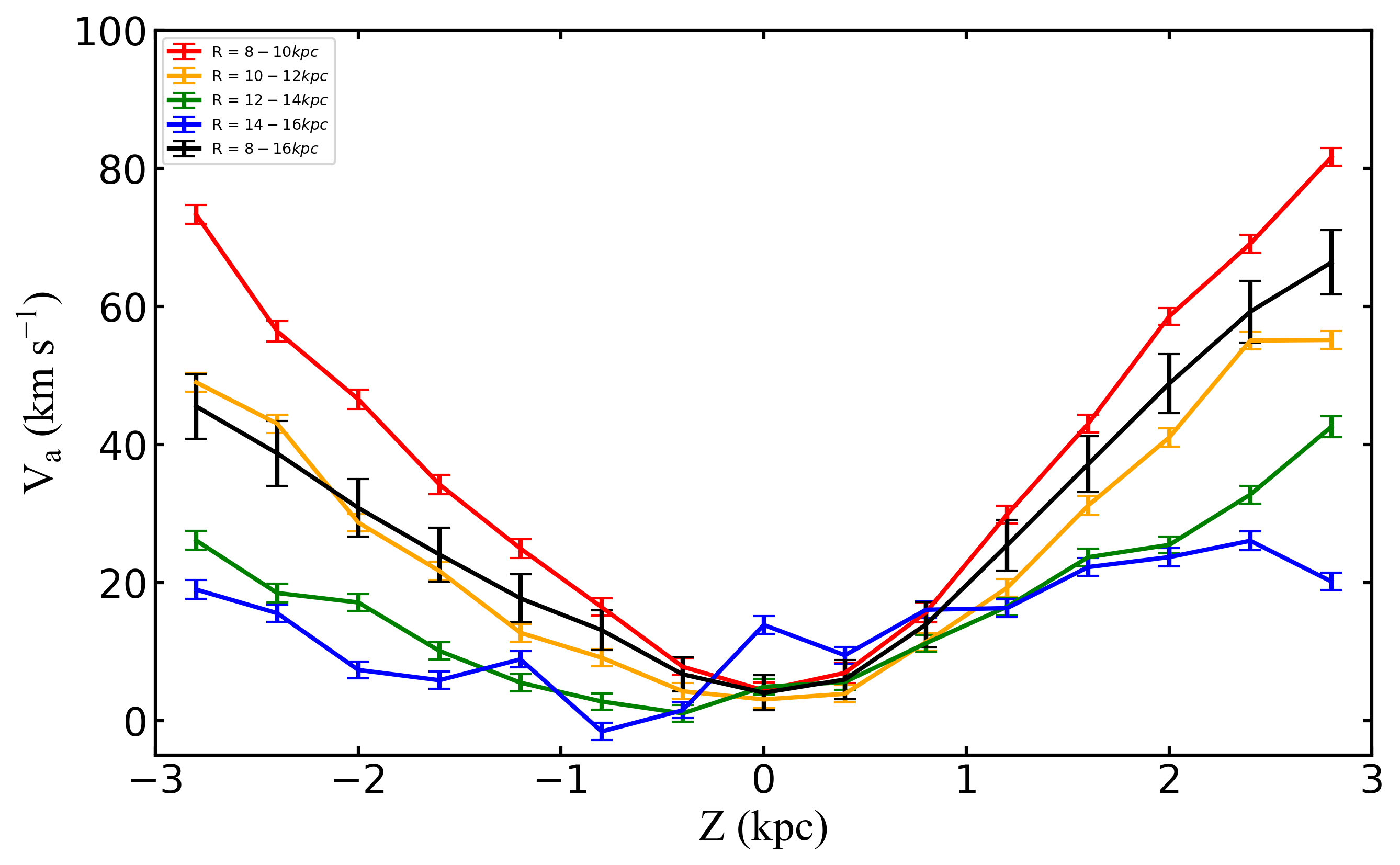}
  \caption{The one-dimensional distribution of asymmetric drift with vertical distance (Z) is presented. Different colors represent sub-samples at various Galactocentric distances (R), with error bars calculated directly from velocity dispersion to provide insight into uncertainty.}
  \label{Z-Va-R}
\end{figure}

As depicted in Figure~\ref{Z-Va-R}, when investigating the pattern of asymmetric drift with the disk's height at different radial distances, we observe an increase in asymmetric drift with vertical distance and a decrease with radial distance from red to blue (black represents the entire sample), consistent with findings in simulation work by \citet{Robin2017}. By computing the median value of the average asymmetric drift on the northern and southern sides of the disk, we observe that the median asymmetric drift on the northern side (20 km/s) is relatively larger than that on the southern side (13 km/s), with errors within 2 km/s. This observation suggests that the ``horn" bends downwards with increasing Galactocentric distance.

\begin{figure}
  \centering
  \includegraphics[width=0.45\textwidth,height=0.35\textwidth]{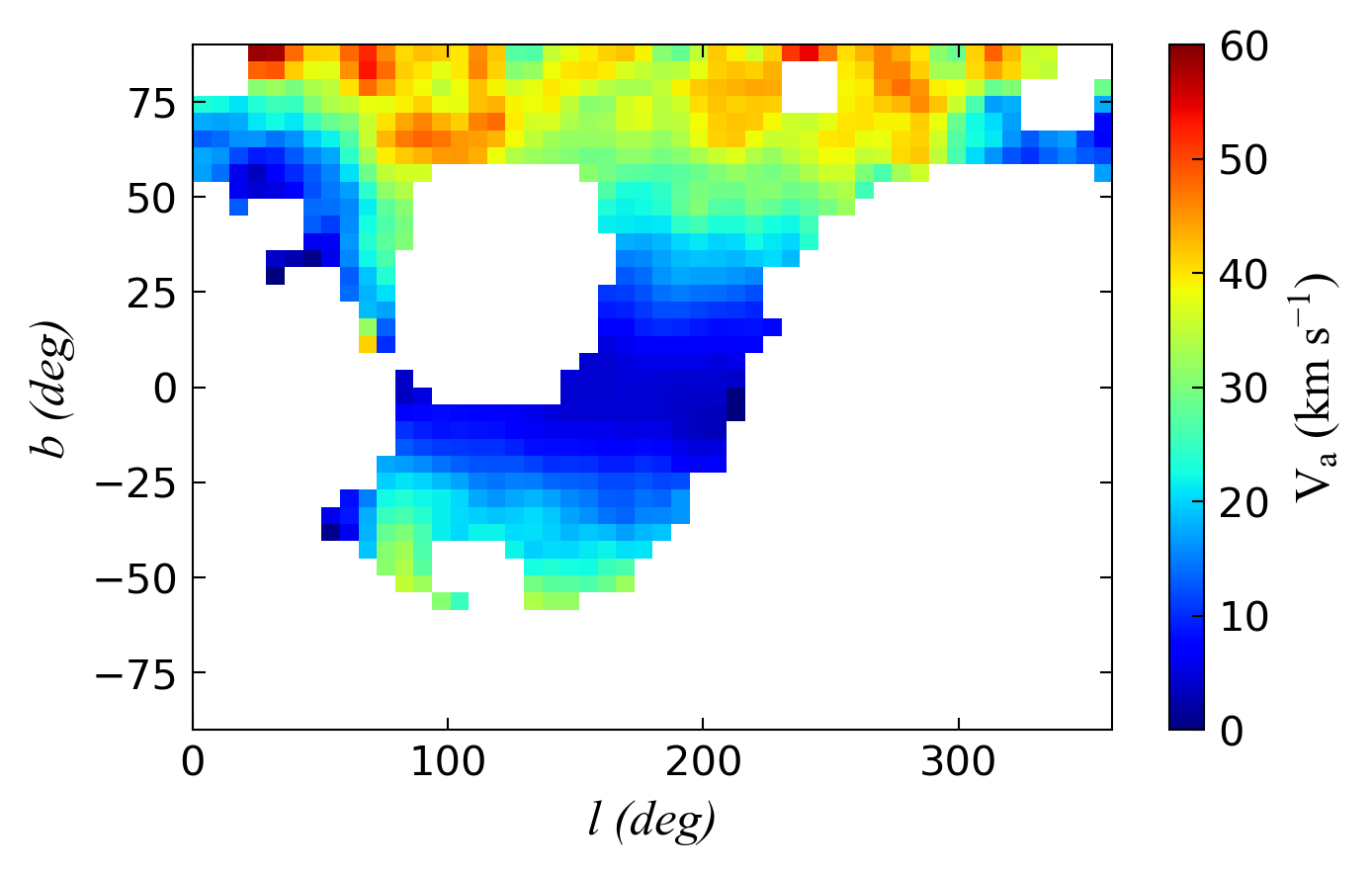}
  \caption{The distribution of asymmetric drift in Galactic celestial coordinates is depicted. It shows that the asymmetric drift is smaller near low Galactic latitudes and larger at higher Galactic latitudes.}
  \label{RA-DEC-VA}
\clearpage
\end{figure}

Moreover, we depict the asymmetric drift in the Galactic latitude-longitude plane in Figure~\ref{RA-DEC-VA}. Similar to the $R-Z$ plane, the asymmetric drift is smaller at lower latitude regions of the Galaxy and larger at higher latitude regions.

\subsection{Asymmetric drift distribution of different age stellar populations}

In this section, we delve into the distribution of asymmetric drift among various stellar age populations, categorized as follows: [0, 2], [2, 4], [4, 6], [6, 8], [8, 10], and [10, 14] Gyr, totaling six populations.

\begin{figure*}
  \centering
  \includegraphics[width=0.9\textwidth,height=0.45\textwidth]{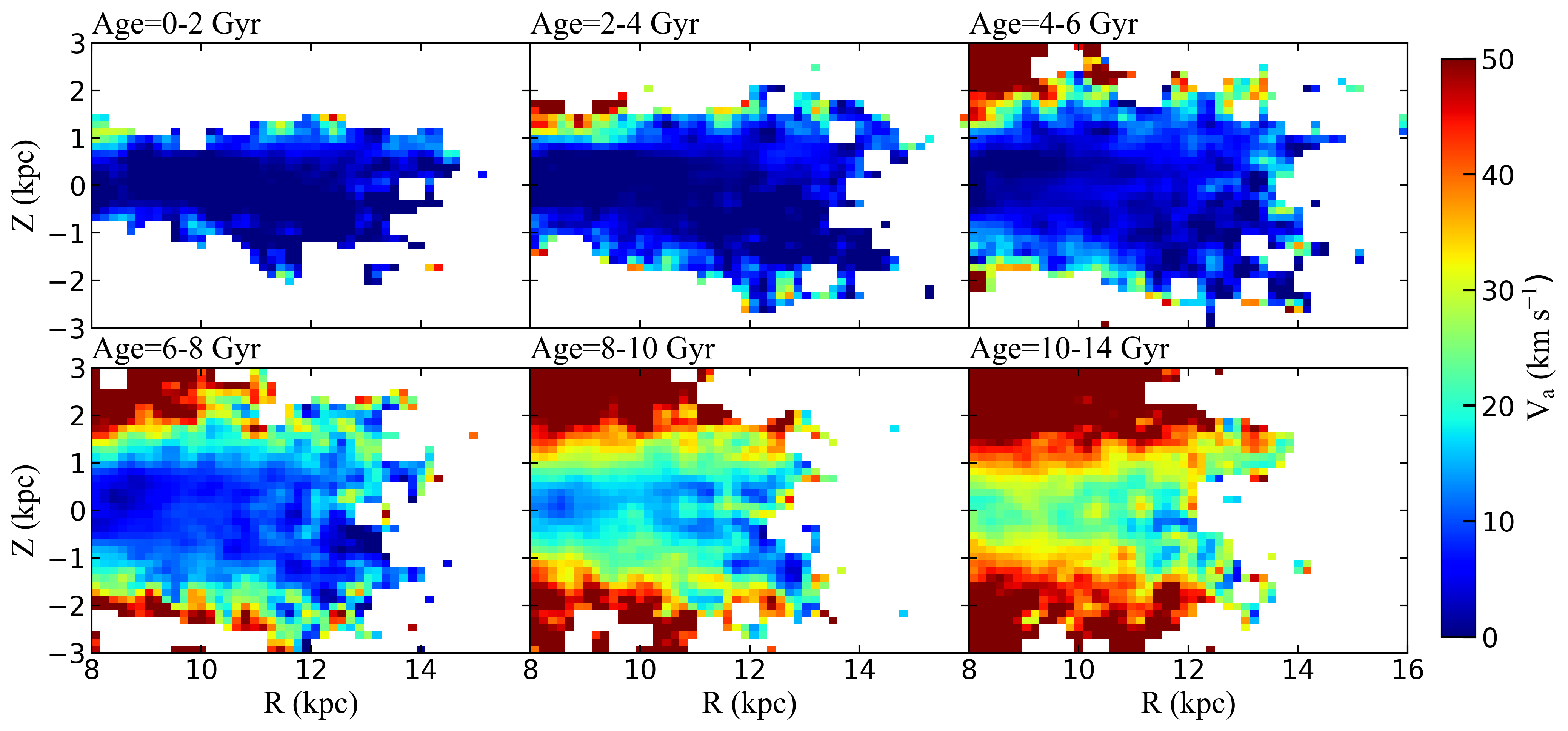}
  \caption{The distribution of asymmetric drift in the $R-Z$ plane is presented for different age populations.}
  \label{R-Z-VAage}
\end{figure*}

Figure~\ref{R-Z-VAage} illustrates the distribution of asymmetric drift in the $R-Z$ plane. Notably, the asymmetric drift for stars younger than 6 Gyr appears relatively smaller, whereas for populations older than 6 Gyr, a larger asymmetric drift is evident, particularly far from the mid-plane. With increasing age, the asymmetric drift on both sides of the disk notably increases. Older stars located farther from the mid-plane exhibit larger asymmetric drift than younger ones near the disk, which aligns with expectations. It's worth noting that previous studies outside the solar neighborhood typically lack age information, hindering the exploration of temporal evolution in asymmetric drift kinematics.

\begin{figure*}
  \centering
  \includegraphics[width=0.9\textwidth,height=0.45\textwidth]{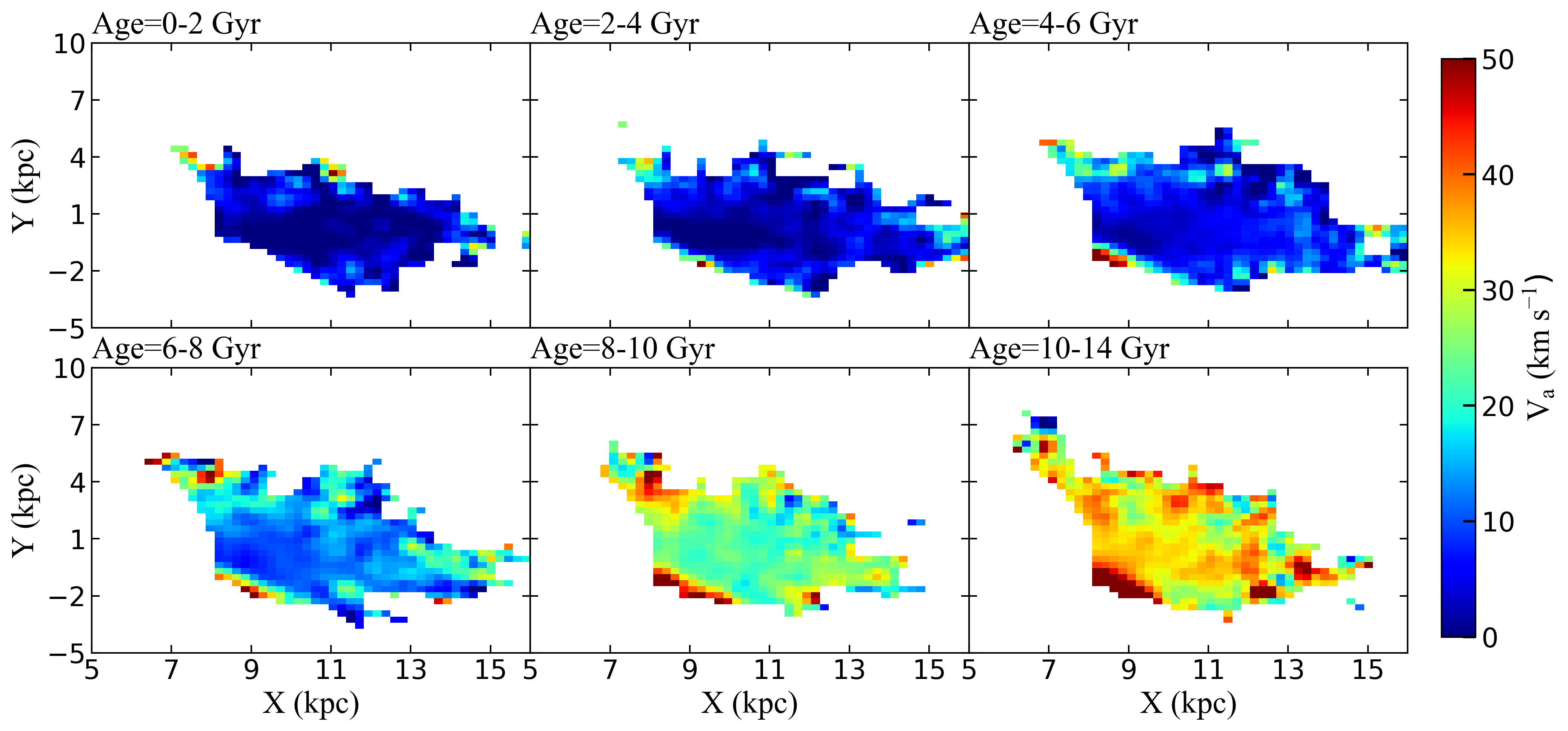}
  \caption{The distribution of asymmetric drift in the $X-Y$ plane is presented for different age populations.}
  \label{X-Y-VA}
\end{figure*}

\begin{figure*}
  \centering \includegraphics[width=0.9\textwidth,height=0.45\textwidth]{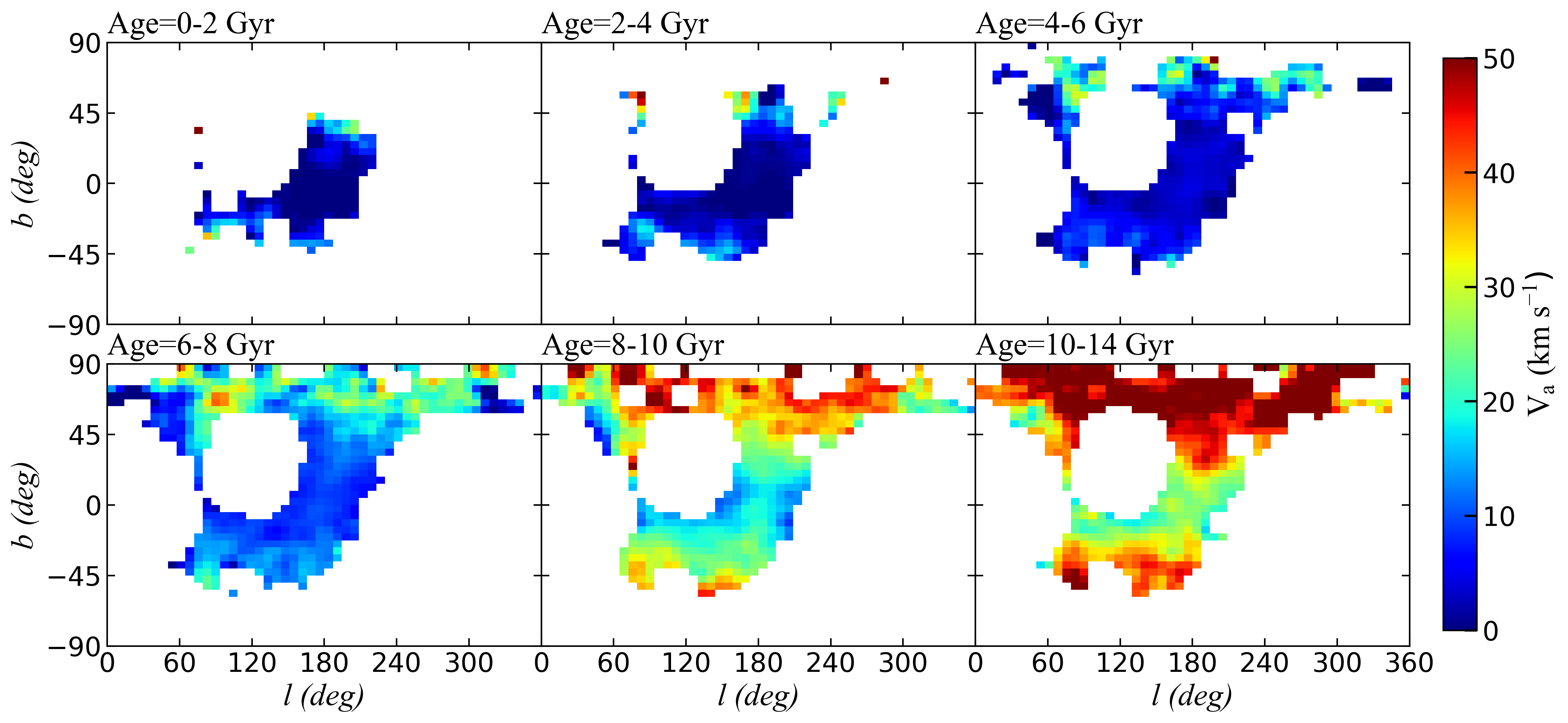}
  \caption{The distribution of asymmetric drift among stellar populations of different ages is depicted in Galactic celestial coordinates.}
  \label{L-B-VA}
\end{figure*}

Similarly, Figure~\ref{X-Y-VA} presents the distribution of asymmetric drift in the X-Y plane for different age bins. It's evident that stars younger than 6 Gyr exhibit relatively smaller asymmetric drift, whereas older stars display significantly larger values. Furthermore, Figure~\ref{L-B-VA} showcases the distribution of asymmetric drift in celestial coordinate systems. Consistently, stars less than 6 Gyr old demonstrate smaller asymmetric drift, while those older than 6 Gyr exhibit larger asymmetric drift. Interestingly, in Figure~\ref{L-B-VA}, two small regions with high asymmetric drift are observed in the 6-8 Gyr range. However, these regions are less visible in other ranges due to sampling and age uncertainty. Considering the R and Z range and referring to our previous work in \citet{wang2018b, wang2020a}, we speculate that these red patches may be attributed to substructures, such as the ``north near" feature located around R = 12 kpc. However, it's important to note that further investigation and analysis are required to confirm these potential substructures. We are currently employing deep learning methods to identify overdensities, substructures, or streams \citep{2024arXiv240105620W}, which may shed more light on this phenomenon in the future.

\begin{figure*}
  \centering \includegraphics[width=0.9\textwidth,height=0.45\textwidth]{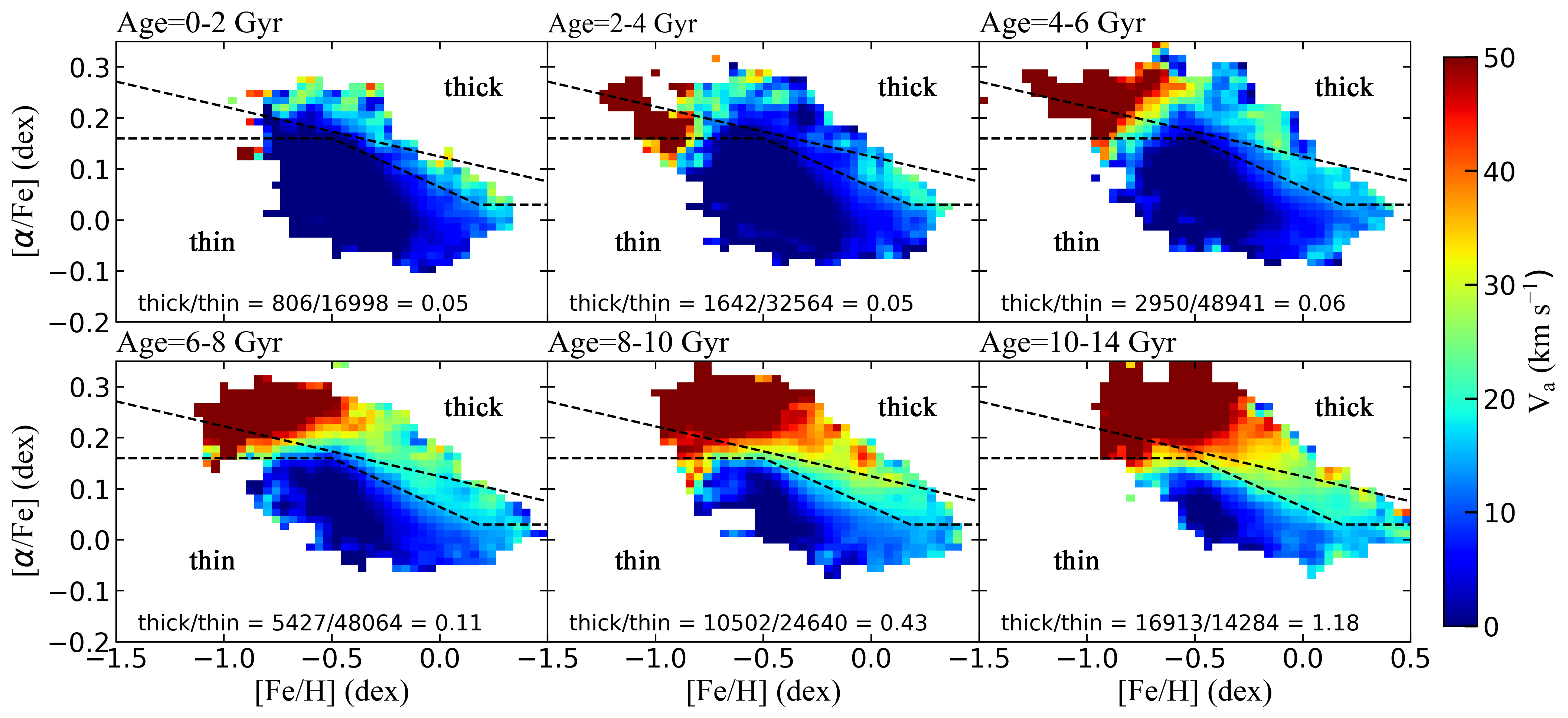}
  \caption{The distribution of asymmetric drift in the [Fe/H] and [$\alpha$/Fe] plane is illustrated for different age populations. The ratio of the number of stars with high $\alpha$ abundance (above the upper dashed line) to those with low [$\alpha$/Fe] abundance (below the lower dashed line) is indicated below each subplot. Time scale picture of star formation for the thin and thick disk can also be seen here.}
  \label{Fe-Alpha-AGE-VA}
\end{figure*}

Figure~\ref{Fe-Alpha-AGE-VA} illustrates the distribution of populations of different ages in chemical space alongside their corresponding asymmetric drift. Notably, high [$\alpha$/Fe] stars exhibit larger asymmetric drift compared to low [$\alpha$/Fe] stars. These distinct clustering features, represented by the blue and red patches, are closely related to the Milky Way's thin and thick disk components. Additionally, we label the fraction of high [$\alpha$/Fe] and low [$\alpha$/Fe] stars in the figure, facilitating further analysis of the thin and thick disk properties. It's observable that as the age increases, the fraction of high [$\alpha$/Fe] stars also increases. This trend can be attributed to the fact that star formation in the thick disk primarily occurred around 8-13 Gyr, involving in-situ gas and the Gaia-Sausage-Enceladus (GSE) system, while the thin disk formation predominantly took place from 6-8 Gyr \citep{2019NatAs...3..932G, 2022arXiv220402989C}. These findings align well with other analyses and figures presented in the current study.

\begin{figure*}
  \centering  \includegraphics[width=0.8\textwidth,height=0.3\textwidth]{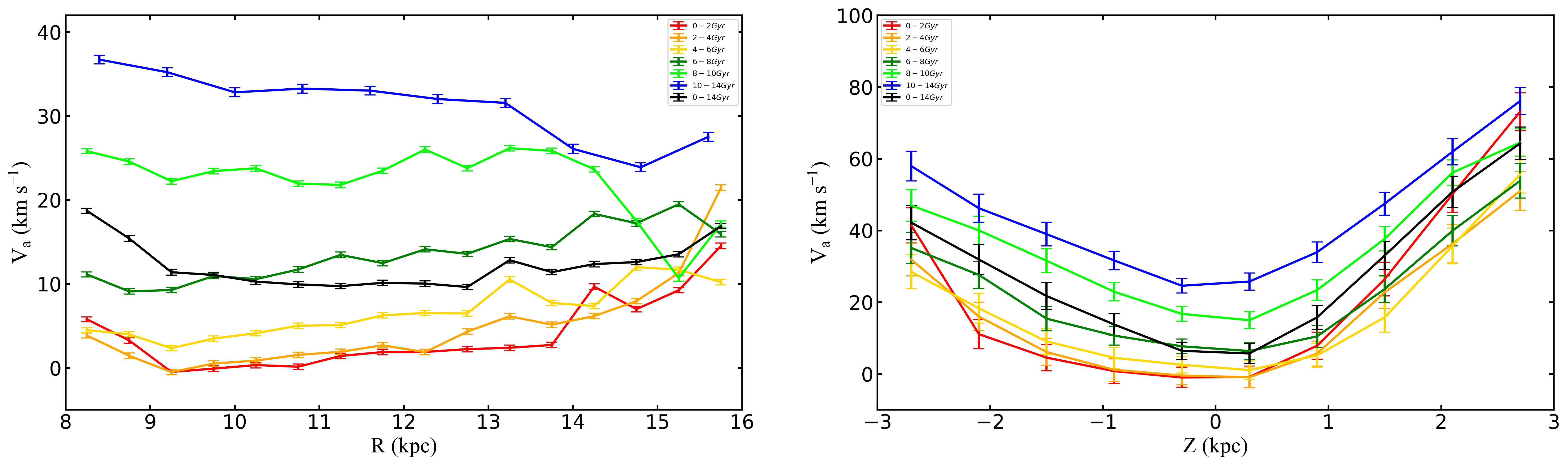}
  \caption{The one-dimensional distribution of asymmetric drift is presented in terms of Galactocentric distance (R) and vertical distance (Z). The black line denotes the entire sample, while solid lines of varying colors represent stellar populations of different ages. Error bars are derived directly from the velocity dispersion.}
  \label{1D velocity}
\end{figure*}

The 1D variation of asymmetric drift with Galactocentric distance and vertical distance across different stellar age populations is depicted in Figure~\ref{1D velocity}. The black line represents the whole sample, while solid lines of various colors represent samples of distinct stellar age populations. Observations reveal that for Galactocentric distances less than 16 kpc, the asymmetric drift exhibits a slight increase for populations younger than 8 Gyr, with a relatively flatter trend within 12 kpc. Conversely, populations older than 8 Gyr display a decreasing trend, albeit with a complex pattern that differs slightly from the modeling results of \citet{Robin2017} within 12 kpc, which were based on mock data and solely utilized high-latitude stars with [Fe/H] to distinguish between the thin and thick disk components. In the case of the youngest populations, a positive gradient with radius is evident, with some regions, such as 13 kpc, displaying fluctuations in the asymmetric drift curve. These fluctuations may arise from non-homogeneous sampling or dynamic perturbations. In the vertical distance diagram, the asymmetric drift curve exhibits a ``U" shape, indicating larger asymmetric drift at greater heights and smaller values near the Galactic disk. This shape and pattern are consistent with previous findings \citep{Golubov2013}.

\subsection{Asymmetric Drift Distribution of Thin and Thick disk}

\begin{figure}
  \centering  
  \includegraphics[width=0.45\textwidth,height=0.35\textwidth]{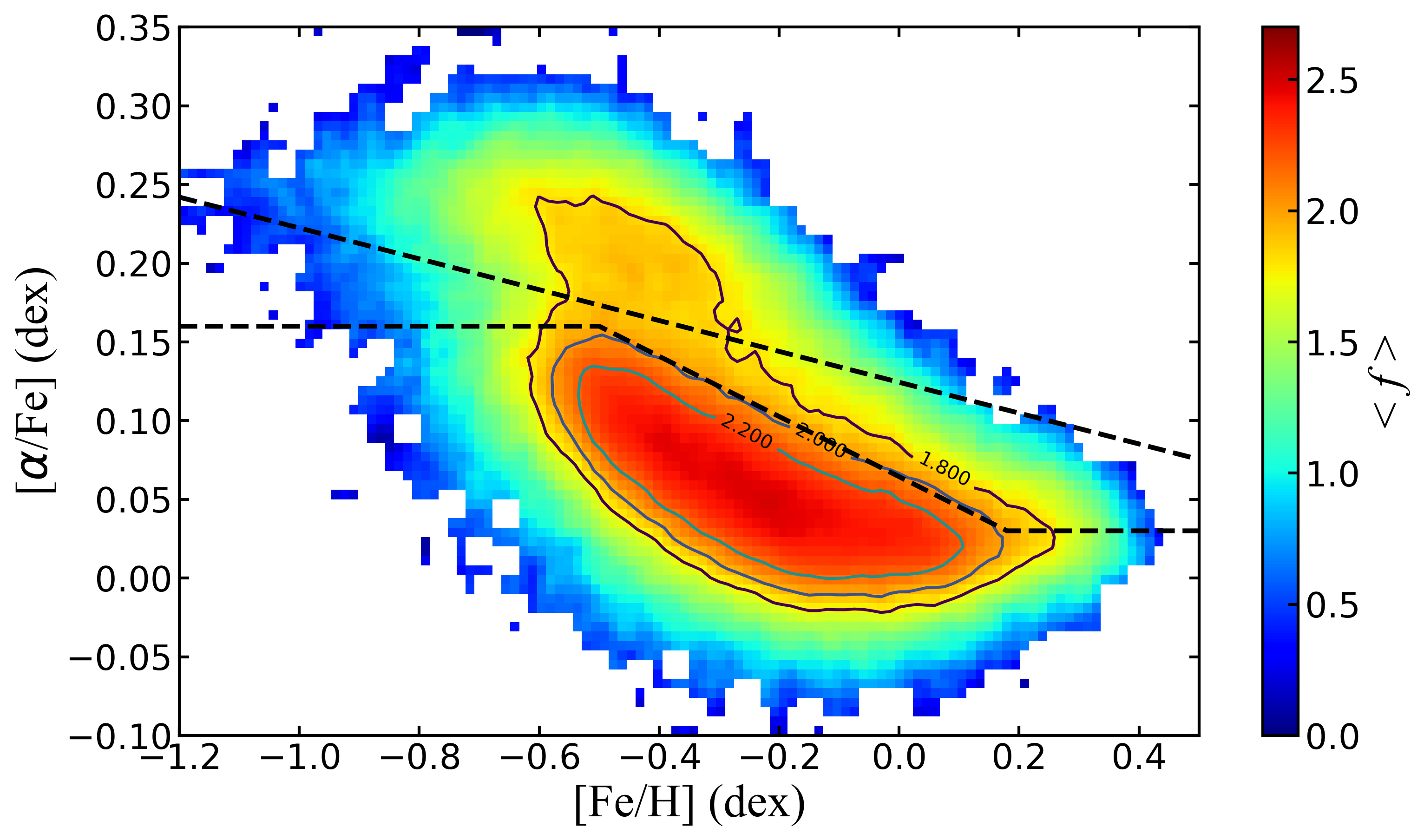}
  \caption{The density distribution of the sample in the [Fe/H] and [$\alpha$/Fe] plane is depicted, with colors representing the logarithmic scale of star counts. Two dashed lines are utilized to distinguish between thin and thick disk stars.}
  \label{FeH-Allpha-f}
\end{figure}

In this section, our primary aim is to investigate the distribution of asymmetric drift in the thin and thick disk components. Figure~\ref{FeH-Allpha-f} illustrates the density distribution of the sample in the metallicity and chemical abundance plane ([Fe/H]-[$\alpha$/Fe]), which we utilize to delineate two distinct populations: thin and thick disk. The thick disk is delineated by the region above the upper dotted line, while the thin disk sample is situated below the lower dotted line.

\begin{figure*}
  \centering
  \includegraphics[width=0.9\textwidth,height=0.35\textwidth]{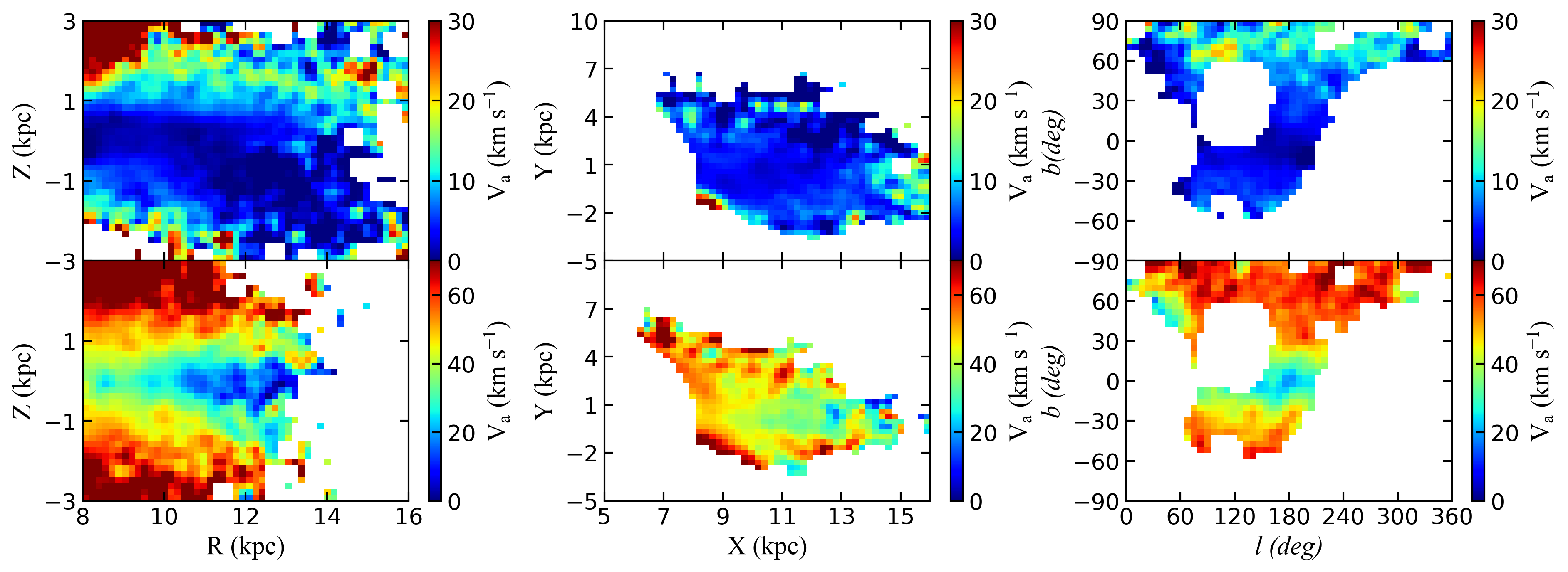}
  \caption{The spatial distribution of thin and thick disk stars is illustrated. In the top panel, the distribution of stars in the $R-Z$, $X-Y$, and $l-b$ planes of the thin disk sample is shown, while the bottom panel displays the distribution of stars in the $R-Z$, $X-Y$, and $l-b$ planes of the thick disk.}
  \label{thin and thick distribution}
\clearpage
\end{figure*}

\begin{figure*}
  \centering
  \includegraphics[width=0.9\textwidth,height=0.55\textwidth]{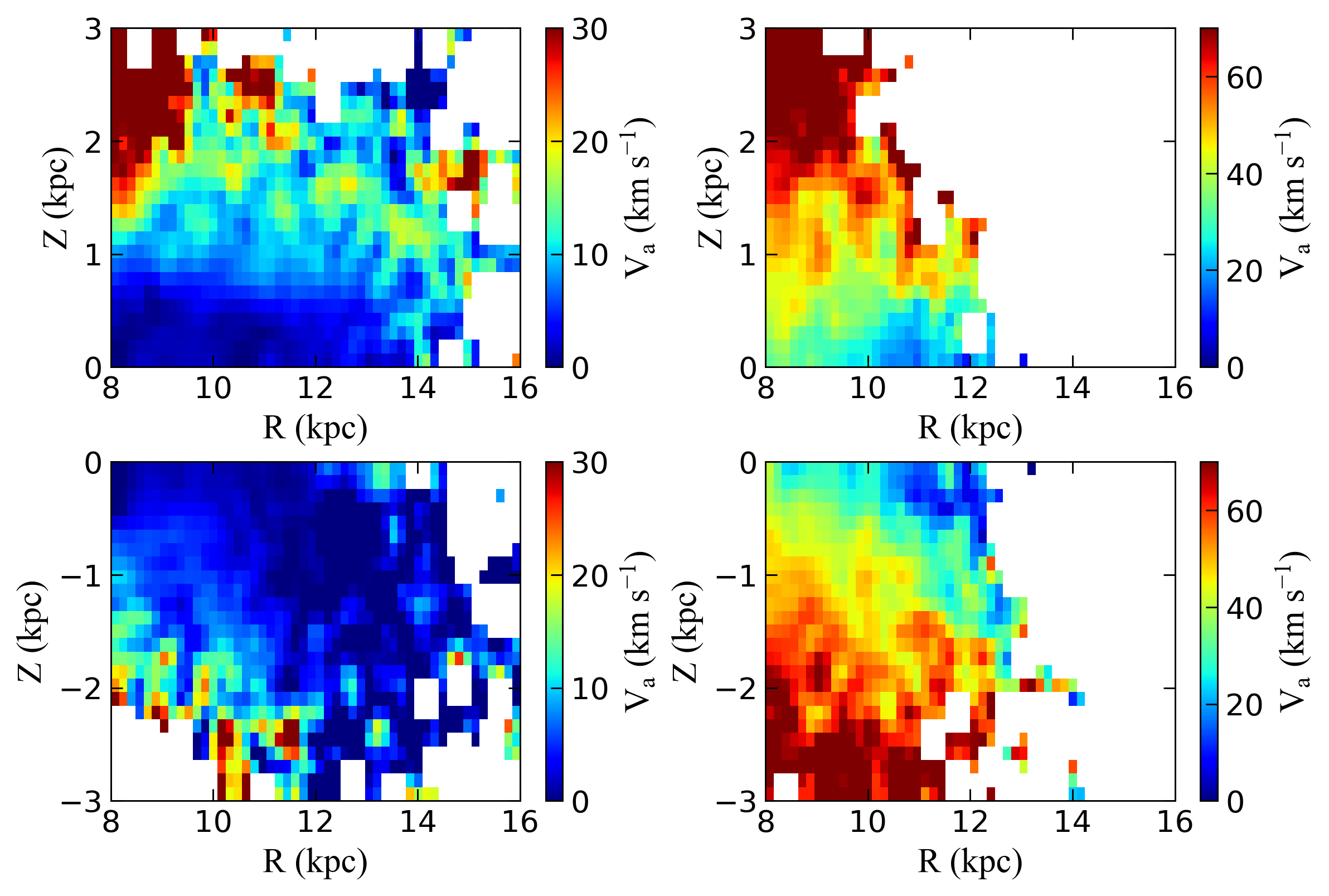}
  \caption{The left panel depicts the distribution of the asymmetric drift of the thin disk on both the north and south sides, while the right panel illustrates the distribution of the asymmetric drift of the thick disk on both sides.}
  \label{N-S thin or thick disk}
\clearpage
\end{figure*}

In the same vein, Figure~\ref{thin and thick distribution} illustrates the distribution of asymmetric drift in the $R-Z$ plane (left), $X-Y$ plane (middle), and Galactic coordinates (right) using samples from the thin and thick disk components (upper panel for thin disk, lower panel for thick disk). Compared to the thin disk, the thick disk exhibits a larger asymmetric drift in spatial distribution. As vertical distance increases, the asymmetric drift rises for both thin and thick disk components, albeit with a notably smaller increase observed for the thin disk. Upon dividing these components into northern and southern sides again, as depicted in Figure~\ref{N-S thin or thick disk}, the variation of asymmetric drift with vertical distance is evident for both the thin (left panel) and thick disk (right panel). Additionally, it is observed that the median value of asymmetric drift for the thick disk (45 km s$^{-1}$) significantly exceeds that of the thin disk (6 km s$^{-1}$), within uncertainties of a few km s$^{-1}$. It is worth noting that our focus lies primarily on qualitative patterns, thus the error margin here is less significant, given our aim to confirm basic knowledge of the disk through population analysis.

\begin{figure*}
  \centering
  \includegraphics[width=0.8\textwidth,height=0.6\textwidth]{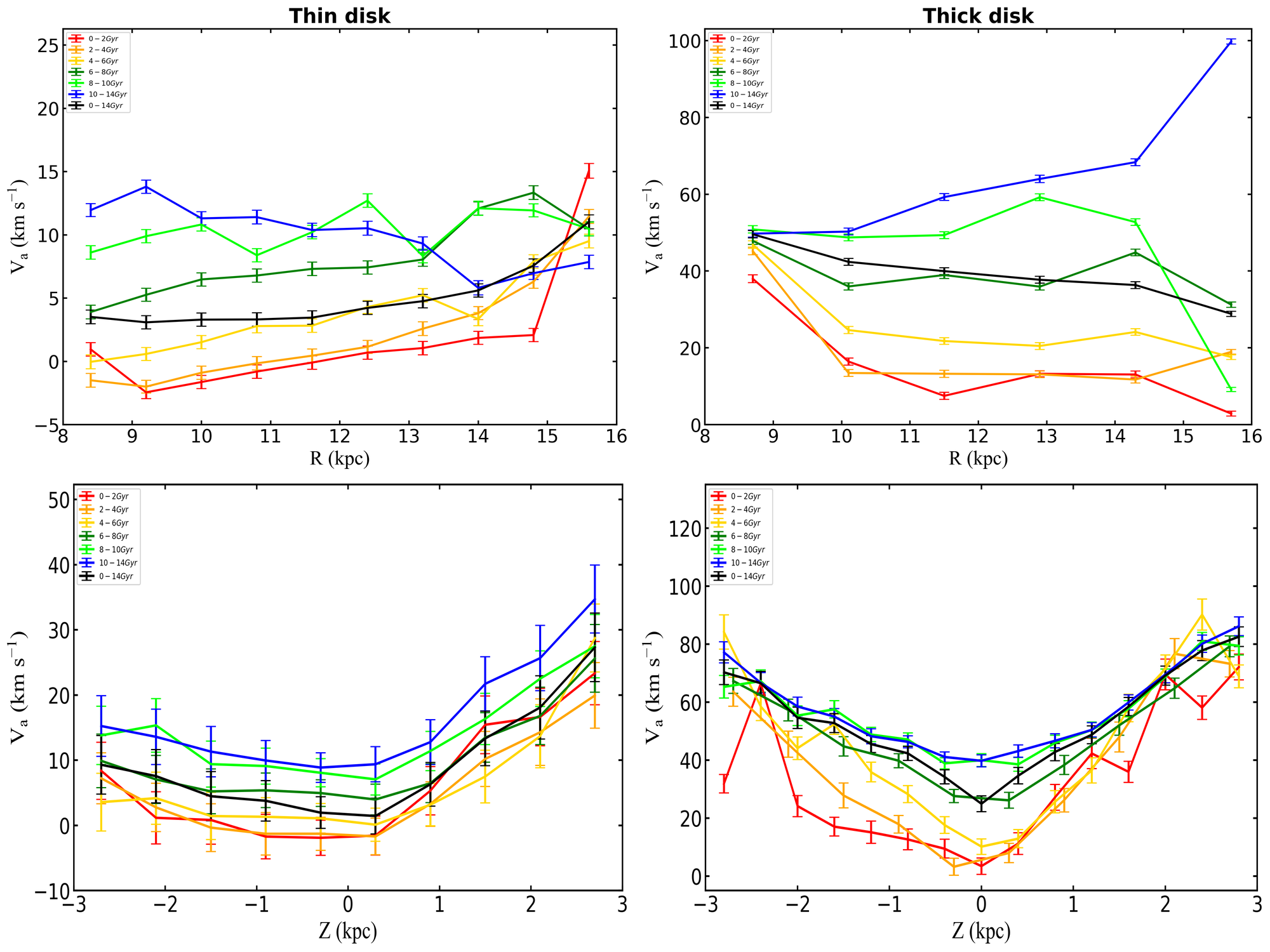}
  \caption{The left two panels display the one-dimensional distribution of the asymmetric drift of the thin disk sample as a function of Galactocentric radial and vertical distance. The right panel illustrates the one-dimensional distribution of the asymmetric drift of the thick disk sample in relation to Galactocentric radial and vertical distance, with error bars representing the dispersion of asymmetric drift values.}
  \label{1D thinthick}
\clearpage
\end{figure*}

Finally, as depicted in Figure~\ref{1D thinthick}, a comparison of the asymmetric drift curves for the thin and thick disk reveals slightly different patterns. The thin disk exhibits an increasing trend for populations younger than 8 Gyr, whereas the thick disk populations display a relatively flatter trend. Beyond 14 kpc, insufficient sampling for some populations may lead to fluctuations. Contrastingly, the asymmetric drift curve of the thin disk with vertical height demonstrates a relatively flatter and smaller trend compared to the thick disk. However, both exhibit a rough ``U" shape pattern, indicating larger asymmetric drift at greater vertical heights near the Galactic disk. It's noteworthy that in the thick disk, the asymmetric drift remains around 50 km s$^{-1}$ for all ages at R = 8 kpc, with drastic changes observed beyond this radius depending on the age. In contrast, for the thin disk, while all populations exhibit different values, they display a similar gradient for populations within 10 Gyr. One possible explanation for these differences could be attributed to sampling variations. Additionally, the distinct kinematic and dynamical properties of the thin and thick disks may play a more significant role. For instance, the thick disk is dynamically hotter than the thin disk. We plan to delve deeper into these dynamics in future work.

\section{discussion}

\subsection{Comparison with other works}

Motivated by the migration history of the disk, \citet{Hayden2018} explored the distribution of asymmetric drift and the orbital properties of metal-rich stars in the solar neighborhood, along with the kinematic properties of stellar populations, using Gaia-ESO survey data. They observed that most [Mg/Fe]-rich stars exhibited significant asymmetric drift, with a velocity 40 km s$^{-1}$ slower compared to lower [Mg/Fe] populations. This correlation between asymmetric drift and chemical composition resonates with our current investigation. Specifically, stars with high [$\alpha$/Fe] exhibit larger asymmetric drift than those with low [$\alpha$/Fe]. In contrast to the Gaia-ESO study, our research delves into exploring the distribution of asymmetric drift across both age and [$\alpha$/Fe] space, particularly focusing on the farther northern outer disk region. By scrutinizing the relationship between median rotational velocity and the square of radial velocity dispersion for three distinct metal-rich populations, \citet{Golubov2013} noted that metal-poor stars displayed smaller asymmetric drift and larger rotational velocities, a finding corroborated by our work (See Figure~\ref{abundance-X}). Our study transcends mere investigation into the distribution of asymmetric drift in chemical space, also revealing the distribution of different age populations.

\begin{figure} \includegraphics[width=0.45\textwidth,height=0.35\textwidth]{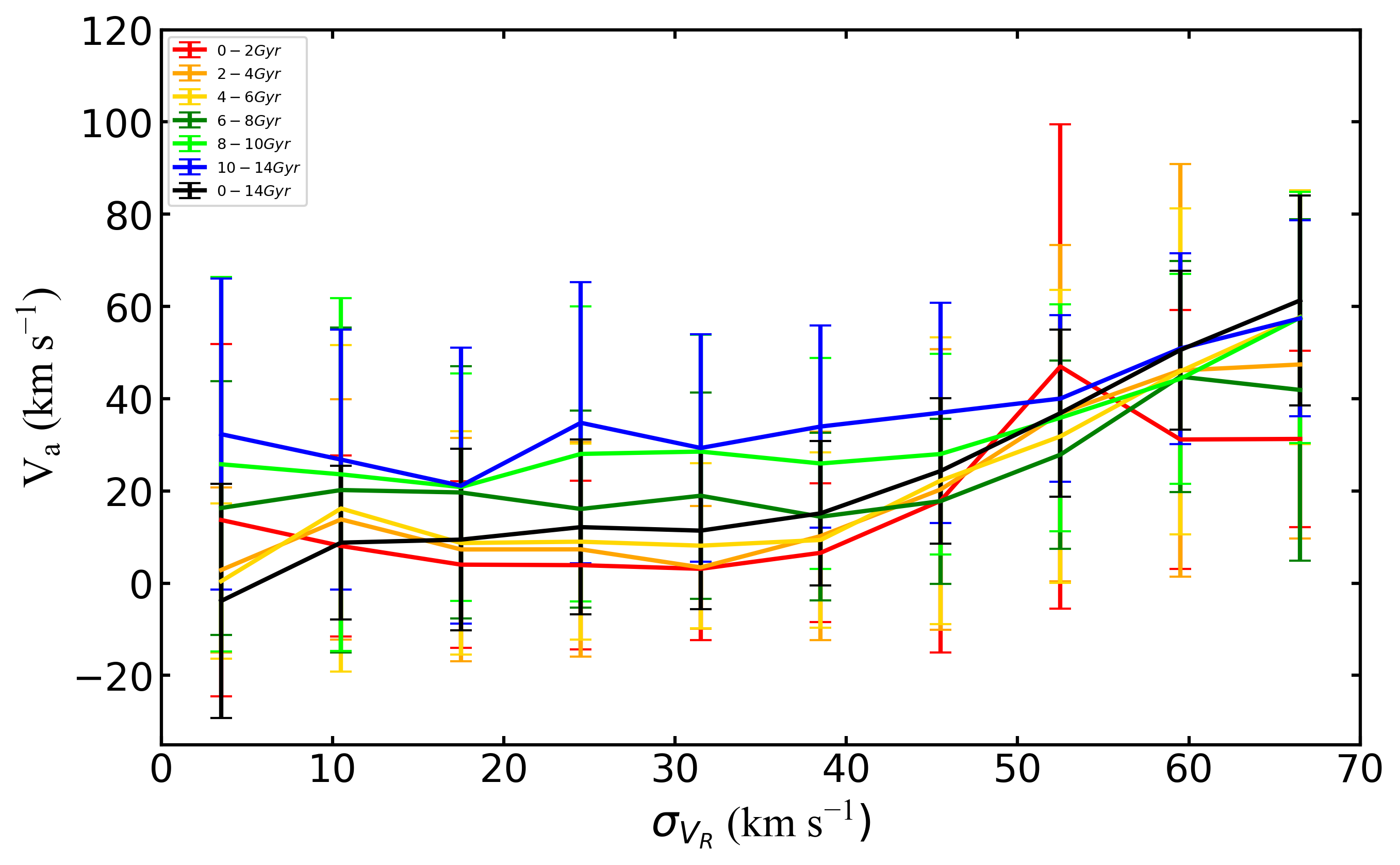}
  \caption{The one-dimensional distribution of asymmetric drift alongside the radial velocity dispersion across different stellar age populations is depicted. The black line corresponds to the entire sample, while solid lines of varying colors represent different stellar age groups.}
  \label{sigmavR-Va}
\end{figure}

\begin{figure*}
  \centering
  \includegraphics[width=0.8\textwidth,height=0.45\textwidth]{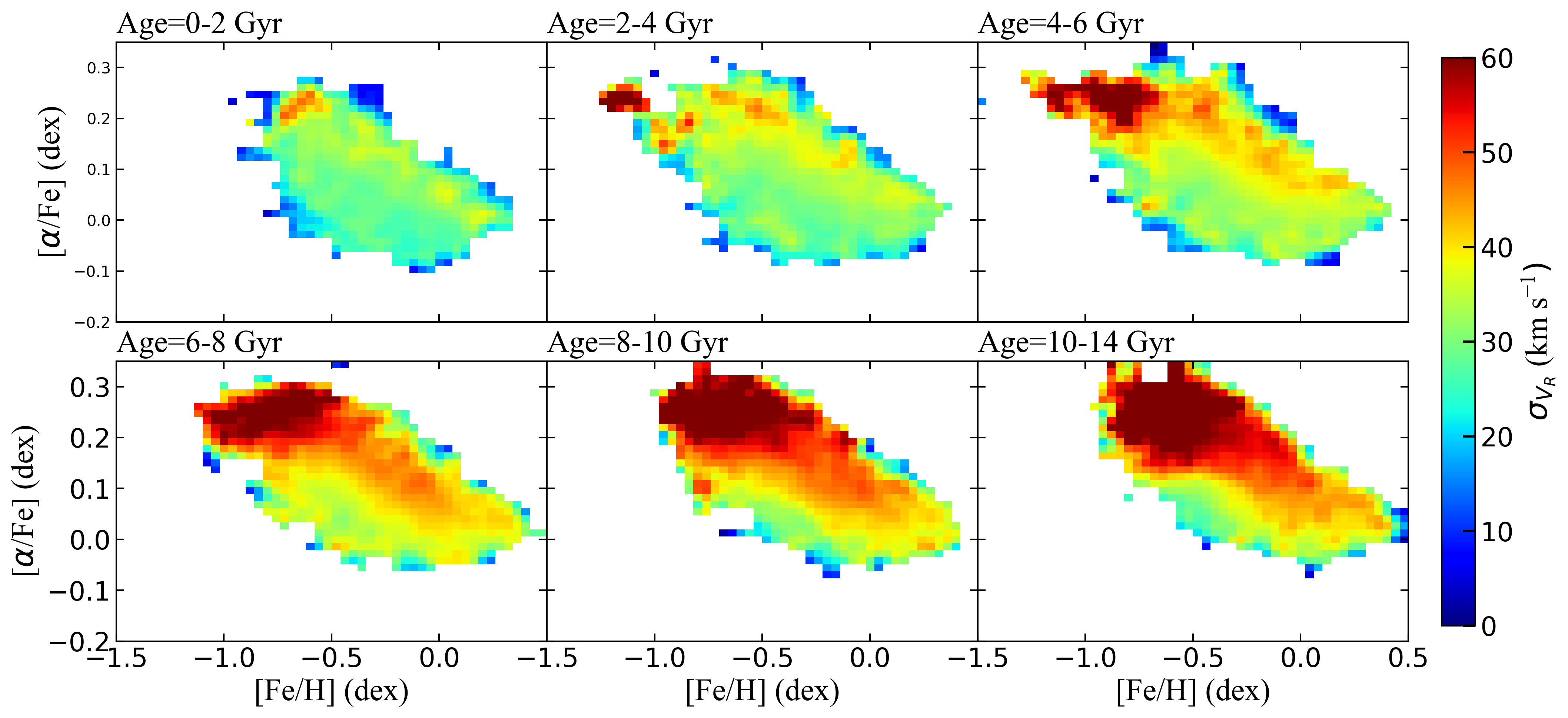}
  \caption{The distribution of radial velocity dispersion in the [Fe/H] and [$\alpha$/Fe] plane for different age populations.}
  \label{Feh-alpha-sigmavR}
\end{figure*}

\begin{figure*}
  \centering
  \includegraphics[width=0.9\textwidth,height=0.95\textwidth]{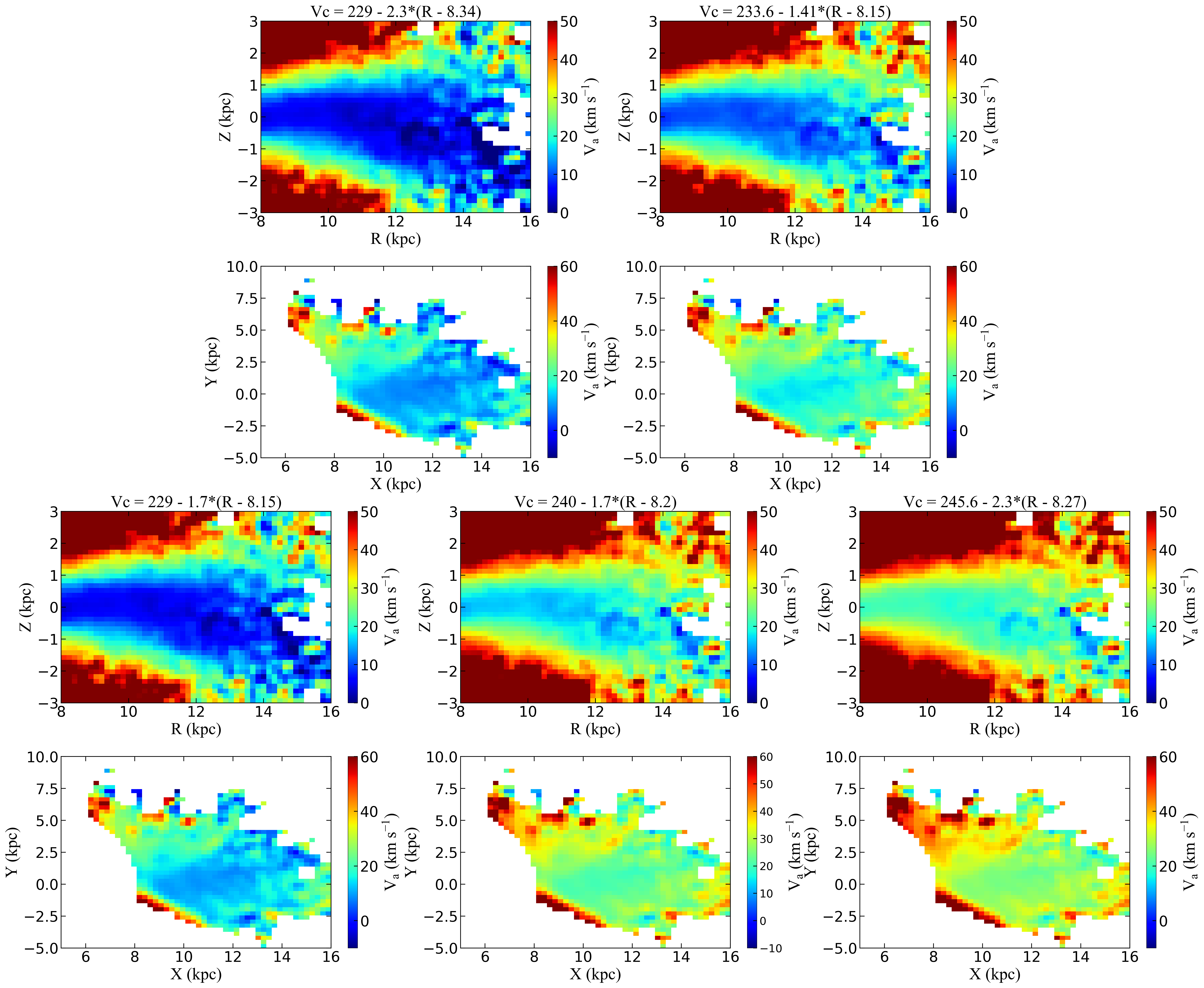}
  \caption{Asymmetric drift diagrams calculated using various rotation curves, solar motions, and solar locations (as described in the text). These diagrams serve as tests to examine the effects of different rotation curves.}
  \label{circular velocity model}
\clearpage
\end{figure*}

\citet{Sysoliatina2018} provided an in-depth analysis of the definition of asymmetric drift, characterizing it as the lag of the tangential velocity of the stellar population concerning the rotation curve. Through iterative calculations starting from the Jeans equation for a stationary and axisymmetric system, they derived a new version of the Str$\ddot{o}$mberg relation and a generalized expression for asymmetric drift. Their findings align with those of \citet{Golubov2013}, indicating that metal-poor stars tend to exhibit smaller asymmetric drift.

Additionally, \citet{Sysoliatina2018} determined a gradient value of 0.98 $\pm$ 1 km s$^{-1}$ kpc$^{-1}$ for the rotational velocity, accounting for the asymmetric drift correction within a Galactocentric distance range of 7 to 10 kpc. While we are extending our observational asymmetric drift (AD) map analysis in spatial and chemical space, it's noteworthy that although purely Gaia datasets can cover this range, comprehensive AD maps in populations within 16 kpc have not been observed to date.

Furthermore, as highlighted by \citet{Golubov2013}, lower metallicity is associated with larger velocity dispersion and radial scale-length, alongside smaller asymmetric drift and faster mean rotation. We further investigate the relationship between asymmetric drift, velocity dispersion, and their distribution on the [Fe/H] and [$\alpha$/Fe] planes (see Figure~\ref{sigmavR-Va} and Figure~\ref{Feh-alpha-sigmavR}). Our analysis reveals that asymmetric drift increases with increasing dispersion, with older stars exhibiting a more pronounced trend. Additionally, the dispersion values, indicative of dynamically cold or hot populations, suggest that the thick disk population (high [$\alpha$/Fe]) is hotter than the thin disk population (low [$\alpha$/Fe]). Generally, younger populations (top three panels) tend to have colder dynamics, while older populations (bottom three panels) exhibit the opposite trend.

Using Gaia DR3 and RAVE DR5 data, \citet{Vieira2022} conducted an investigation into the kinematic properties of the thin and thick disk in the Solar neighborhood. They observed that the median rotational velocities of both the thin and thick disk exhibit a lag behind the assumed velocities of the local standard of rest. Specifically, they noted that the thin disk lags the local standard of rest by 5 to 8 km s$^{-1}$, while the median rotation velocity of the thick disk in the solar neighborhood lags the local standard of rest by 20 km s$^{-1}$. This lag is attributed to the phenomenon of asymmetric drift. \citet{Vieira2022} identified two primary reasons for this asymmetric drift phenomenon. Firstly, it arises from the disparity between the gravitational and centrifugal forces caused by the non-zero component of the velocity dispersion of the system. Secondly, it results from the mixing of the thin and thick disk components. Our findings in this study support these observations, particularly highlighting the significantly larger asymmetric drift of the thick disk compared to the thin disk across various population details. In summary, our current results corroborate significant aspects of the overall trend and key arguments put forth by previous studies, providing additional insights into the detailed population dynamics and extending observations to farther distances \citep{Pasetto2012, Guiglion2015, Wojno2016, Robin2017}.

Some modelling studies of Milky-Way-like galaxies have predicted a systematic variation of the asymmetric drift with age and metallicity \citep{Schonrich2010,Loebman2011}. Notably, \citet{Lee2011b} demonstrated, using the SEGUE G dwarf sample, that the asymmetric drift in the thin disk decreases with decreasing metallicity. This finding contrasts with the naive expectation of local evolution models. From a qualitative standpoint, our observational results align with these previous studies and simulations. However, our work benefits from a broader coverage and more detailed population information. This expanded dataset enables us to gain a deeper understanding of disk kinematics and dynamics in the outer regions of the Milky Way.

\subsection{Different rotation curves comparisons}

\citet{Eilers2019}, \citet{Mroz2019}, and \citet{wang2023a} reported rotation curve gradients of $-$1.7 km s$^{-1}$ kpc$^{-1}$, $-$1.41 km s$^{-1}$ kpc$^{-1}$, and $-$2.3 km s$^{-1}$ kpc$^{-1}$, respectively. These values are largely consistent within the scope of this study, albeit with slightly different slopes. Variations in the solar radius, motions, and the local standard of rest (LSR) among different research groups \citep{Bland2016, Khoperskov2022, Schonrich2012} may contribute to these discrepancies.

Figure~ \ref{circular velocity model} presents the distribution of asymmetric drift in the $R-Z$ and $X-Y$ planes based on different rotation curve models, solar locations, motions, and LSR values, providing a comparative analysis. While different circular velocity models and solar motions influence the patterns and values of the asymmetric drift, the overall qualitative trends remain similar across variations. It is evident that the main systematic effects arise from differences in the circular velocity at the Sun. This study primarily focuses on elucidating the statistical laws governing the asymmetric drift in various populations beyond the Solar neighborhood. 

\section{CONCLUSIONS} 

In this study, we conduct a cross-match between LAMOST DR4 red giant branch (RGB) stars, characterized by their age and abundance, and Gaia DR3 proper motion data. Our aim is to investigate the distribution of asymmetric drift in the outer regions of the Galaxy across different stellar populations. Our analysis reveals that the median asymmetric drift for the entire sample within the range ($R$ = [8, 16] kpc, $Z$ = [$-$3, 3] kpc) is 16 km s$^{-1}$, while the value near the Sun is 6 km s$^{-1}$. Examining the asymmetric drift distribution in the $R-Z$ plane, we observe a distinctive ``horn" shape pattern. Specifically, the asymmetric drift varies with the radial distance from the Galactic center, with the median asymmetric drift higher on the northern side of the Galactic disk (20 km s$^{-1}$) compared to the southern side (13 km s$^{-1}$).

Our investigation also uncovers a noticeable trend of increasing asymmetric drift with vertical distance, which varies across different stellar populations. Analysis based on stellar age demonstrates that younger stars exhibit smaller asymmetric drift values, whereas older stars display larger asymmetric drift values, consistent with observed velocity dispersion patterns. The one-dimensional asymmetric drift diagram further supports these findings, illustrating a gradual increase in asymmetric drift with vertical distance and resulting in a characteristic ``U" shaped pattern. Furthermore, our analysis in chemical space reveals that stars with high [$\alpha$/Fe] ratios tend to exhibit larger asymmetric drift values compared to those with low [$\alpha$/Fe] ratios. Additionally, we find that the thick disk population generally exhibits a larger asymmetric drift, reaching up to 45 km s$^{-1}$. In summary, our study provides a more comprehensive and extensive map of the asymmetric drift in the Milky Way disk populations within the range of 8$-$16 kpc, contributing valuable insights to the astronomical community.

Indeed, our analysis relies on Eq. (\ref{model2}), which assumes a fixed rotation curve and derives variations in asymmetric drift based on this assumption. However, it's worth noting that the rotation curve could potentially vary with both height above the Galactic plane ($Z$) and stellar age, introducing a degeneracy in our solutions. Unfortunately, the precise laws governing these dependencies remain unclear at present. Given this uncertainty, we have limited our analysis to the $Z$ = $-$3 to 3 kpc disk region, where the rotation curve slope is known. While this approach may not fully disentangle the degeneracy, it provides valuable insights into the asymmetric drift within this specific region. Future studies may aim to refine our understanding of the rotation curve's dependence on $Z$ and stellar age, thus enhancing our ability to accurately model the asymmetric drift across the Galaxy. 

Moreover, the exploration of different solar motions and rotation curve tests has provided some mitigation against the influences of degeneracy. However, it's essential to acknowledge that our contribution primarily focuses on qualitative analysis, a dimension that has been somewhat neglected in prior research. Again, not only the intricate interplay between rotation curve and age dependencies remains inadequately understood within the scientific community, but also the relationship between the rotation curve and height ($Z$) dependence, potentially influenced by asymmetric effects, may indeed merit further and more comprehensive investigation in future studies. Such endeavors could yield valuable insights into the dynamics of the Galactic disk and enhance our understanding of its evolution over time.

The non-equilibrium state and intricate dynamical perturbations within the Milky Way are fundamental factors contributing to the origins of asymmetric drift, a crucial parameter pivotal for unraveling the Galaxy's dynamical history and evolutionary trajectory. It exhibits physical correlations with various properties, including 6D phase space characteristics, radial scale length, gravitational potential, and stellar populations, among others. In our future endeavors, we aim to employ a diverse array of methodologies to comprehensively measure the asymmetric drift. Additionally, we plan to integrate simulations into our investigations to further illuminate the complexity of asymmetric drift, thereby enhancing our understanding of solar motions, kinematics, dynamics, and the rotation curve of the Milky Way disk.

 \section*{Acknowledgements}

We would like to thank the anonymous referee for his/her very helpful and insightful comments. This work is supported by the National Key R\&D Program of China (Nos. 2021YFA1600401 and 2021YFA1600400), the National Natural Science Foundation of China (NSFC) under grant 12173028, the Chinese Space Station Telescope project: CMS-CSST-2021-A10, the Sichuan Youth Science and Technology Innovation Research Team ( Grant No. 21CXTD0038) and  the Innovation Team Funds of China West Normal (No. KCXTD2022-6). 

The Guo Shou Jing Telescope (the Large Sky Area Multi-Object Firber Spectroscopic Telescope, LAMOST) is a National Major Scientific Project built by the Chinese Academy of Sciences. Funding for the project has been provided by the National Development and Reform Commission. LAMOST is operated and managed by National Astronomical Observatories, Chinese Academy of Sciences. This work has also made use of data from the European Space Agency (ESA) mission {\it Gaia} (\url{https://www.cosmos.esa.int/gaia}), processed by the {\it Gaia} Data Processing and Analysis Consortium (DPAC, \url{https://www.cosmos.esa.int/web/gaia/dpac/consortium}). Funding for the DPAC has been provided by national institutions, in particular the institutions participating in the {\it Gaia} Multilateral Agreement.

%%%%%%%%%%%%%%%%%%%%%%%%%%%%%%%%%%%%%%%%%%%%%%%%%%

%%%%%%%%%%%%%%%%%%%% REFERENCES %%%%%%%%%%%%%%%%%%

% The best way to enter references is to use BibTeX:
%\vspace*{-\baselineskip}
%\bibliographystyle{mnras}
%\bibliography{reference.bbl} % if your bibtex file is called example.bib

% Alternatively you could enter them by hand, like this:
% This method is tedious and prone to error if you have lots of references

% Don't change these lines
%\bsp	% typesetting comment
\label{lastpage}
\end{document}